\documentclass[usenatbib,useAMS]{mn2e}

\usepackage{times}
\usepackage{epsfig}
\usepackage{color}

\begin{document}




\title[Compact objects from binary self-lensing]
{Compact object detection in self-lensing binary systems with a main-sequence star}

\author[S. Rahvar, A. Mehrabi and M. Dominik]
{S. Rahvar$^{1,2}$\thanks{rahvar@sharif.edu}, A. Mehrabi$^1$ and
M. Dominik$^3$\thanks{Royal Society University Research Fellow}\\
$^1$Department of Physics, Sharif University of Technology,
P.O.~Box 11365--9161, Tehran, Iran\\
$^2$School of Astronomy, IPM (Institute for Studies in Theoretical Physics and Mathematics), P.O. Box 19395-5531, Tehran, Iran \\
$^{3}$SUPA, University of St Andrews, School of Physics \&
Astronomy, North Haugh, St Andrews, KY16 9SS, United Kingdom }

\maketitle

\begin{abstract}

Detecting compact objects such as black holes, white dwarfs, strange
(Quark) stars and neutron stars by means of their gravitational
lensing effect on an observed companion in a binary system has
already been suggested almost four decades ago. However, these
predictions were made even before the first observations of
gravitational lensing, whereas nowadays gravitational microlensing
surveys towards the Galactic bulge yield almost 1000 events per year
where one star magnifies the light of a more distant one. With a
specific view on those experiments, we therefore carrry out
simulations to assess the prospects for detection of the transient
periodic magnification of the companion star, which lasts typically
only a few hours binaries involving a main-sequence star. We find
that the effect is practically independent of the distance of the
binary system from the observer, but a limit to its detectability is
given by the achievability of dense monitoring with the required
photometric accuracy. In sharp contrast to earlier expectations by
other authors, we find that main-sequence stars are not
substantially less favourable targets to observe this effect than
white dwarfs, not only because of a better achievable photometry on
the much brighter targets, but even more due to the fact that there
are $\ga 10^4$ times as many objects that can be monitored. The
requirement of an almost edge-on orbit leads to a probability of the
order of $3 \times 10^{-4}$ for spotting the signature of an
existing compact object in a binary system with this technique.
Assuming an abundance of such systems about 0.4 per cent, a
high-cadence monitoring every 15~min with 5 per cent photometric
accuracy would deliver a signal rate per target star of $\gamma \sim
4 \times 10^{-7}~\mbox{yr}^{-1}$ at a recurrence period of about 6
months. With microlensing surveys having demonstrated the capability
to monitor about $2 \times 10^{8}$ stars, one is therefore provided
with the chance to detect roughly semi-annually recurring
self-lensing signals from several compact compacts in a binary
system. These must not be mistaken for similar signatures that arise
from isolated planetary-mass objects that act as gravitational lens
on a background star. If the photometric accuracy was pushed down to
0.3 per cent, 10 times as many signals would become detectable.

\end{abstract}

\begin{keywords}

black hole -- binary stars -- gravitational lensing

\end{keywords}

\section{Introduction}
Despite the successful observation of the bending of light by the
Sun \citep{Dyson:eclipse}, following the suggestion by
\citet{Ein11}, it required many decades of advance in technology for
enabling the detection of this effect for another star, given that
"there is no great chance of observing this phenomenon"
\citep{Ein36}. Only following the call by \citet{Pac86} to apply
'gravitational microlensing' to measure the abundance of potential
MACHOs (Massive Compact Halo Astrophysical Objects) in the Galactic
halo, the first related experiments were carried out. In fact, a
decade of observations of the Large and Small Magellanic Clouds now
reveals that there are not enough MACHOs in the Galactic halo to
account for the observed  flat rotation curve for the Galactic disk
\citep{mil01,Popowski,mon09}. The gravitational microlensing effect has
evolved into an important astrophysical tool for not only studying
stellar atmospheres \citep[e.g.\ ][]{alb99,afo01,GouldRev,abe03}, but also to
study populations of extra-solar planets
\citep{MP91,GL92,Do:planetreview}.

In this work, we assess the suggestion to detect Compact Objects
(CO), namely black holes (BH), strange (Quark) stars (QS) and
neutron stars (NS), by means of their gravitational bending of light
received from an observed star that forms a binary system together
with the Compact Object in the context of current experiments. The
lens action within a binary system of stars or stellar remnants has
been discussed in great detail by \citet{Mae73}. This effect shares
many characteristics with the meanwhile common gravitational
microlensing events where a foreground star magnifies the light of
an unrelated background star, which get aligned on the sky with
respect to the observer just by chance. However, the typical
duration of the transient brightening is substantially shorter, of
the order of a few hours, and the signal repeats periodically
(albeit with periods that can be as large as decades). \citet{Mae73}
moreover found that the smaller the radius of the source star, the
larger is the lens effect and its probability of occurence. As a
consequence, main-sequence stars (MS) were considered unfavourable
candidates as compared to white dwarfs, where however the prospects
for MS-BH pairs are substantially better than for MS-NS and MS-WD
pairs. As a consequence, \citet{BT} have more recently evaluated the
detectability of compact objects in a binary system with an observed
white dwarf due to gravitational lensing, and in particular looked
at the prospects for observing this effect in the Sloan Digital Sky
Survey (SDSS), while not considering main-sequence source stars.

However, the chances of success in both cases depend on a number of
various factors. First, there is the existing number of respective
pairs of binary systems, on which we are currently forced to rely on
the best available understanding of stellar evolution. Observations
of star forming regions show that 70 to 90 per cent of stars form in
the clusters and almost two out of three stars reside in binary
systems \citep{mat}. Models of stellar evolution predict that 0.4
per cent of the binary systems will see one of companions turning
into a compact object \citep{hur00,bel02}, whereas 0.2 per cent of
stars end up in a binary system composed of two compact objects.
Second, the probability for a signature to be ongoing at any time is
given by the product of the probability for the monitored target to
show a signal and the ratio between the signal duration and the
orbital period. Third, the number of suitable targets that can be
monitored plays a crucial rule, and fourth and finally, it cannot be
neglected that high-precision photometry on main-sequence stars as
far as the Galactic bulge is possible, whereas such an opportunity
does not arise for the much fainter white dwarfs.

Gravitational lensing of a star gravitationally bound to a compact object
has also been proposed by \citet{cam73} as an interpretation
of the Weber experiment \citep{web70} for the
gravitational radiation from the center of Galaxy, where they used the
optical approach for calculating the lensing effect in a
Schwarzschild  metric when the source star is aligned with the
massive black hole of the Galaxy and the observer. In the optical
approach, the variation of light bundle along the null geodesic
describes the intensity of the light. In the extension of this work,
\cite{cun73} obtained the gravitational lensing of a source star
rotating around a maximally Kerr metric. The main physical
difference between the lensing in the work by \cite{cam73} and
eclipsing microlensing proposed in this work is that in the former
case the source star is orbiting around the black hole with the
orbital size in the order of Schwarzschild radius while in later
case the source is located in the order of the Astronomical Unit. In
this case, the line between the source-lens and the optical axis
(line connecting lens to the observer) is small \citep{bozz05}.

In contrast to \citet{BT}, we focus on the self-lensing within
binaries that are composed of a compact object and an observed
main-sequence star, and on the observability of this effect with
current or upcoming microlensing monitoring efforts.

In Sect.~\ref{Binary}, we discuss the arising binary self-lensing
light curves, and subsequently evaluate the detection probability of
such signals using strategies similar to ongoing microlensing
efforts in Sect.~\ref{montecarlo} by means of Monte-Carlo
simulations. We briefly discuss the extraction of parameters from
the observed data in Sect.~\ref{deg}, before we finally summarize
our conclusions in Sect.~\ref{conclusion}.

\section{Self-microlensing within binary systems}
\label{Binary}

\begin{figure}
\begin{center}
\includegraphics[height=3.5cm,width=6.5cm]{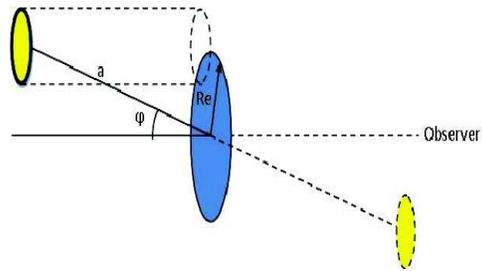}
\caption{\label{fig1} Geometrical configuration of lens and source
in a binary system. The horizontal line represents the observer-lens
line of sight. The binary system with the observer is shown from the
side, and $\varphi$ denotes the inclination angle of the binary
system with respect to observer-lens line of sight. For simplicity,
we assume circular orbits with a radius $a$.} \label{fig:config}
\end{center}
\end{figure}

As illustrated in Fig.~\ref{fig:config}, the self-lensing binary
system involving the compact object is characterised by its
inclination angle $\varphi$ with respect to the observer-lens axis
(the lens being the compact object), the orbital radius $a$
(assuming circular orbits for simplicity), and the Einstein radius
\begin{equation}
R_\rmn{E} = \sqrt{2\,R_\mathrm{S}\,a}\,,
\end{equation}
where
\begin{equation}
R_\rmn{S} = \frac{2GM}{c^2}\,,
\end{equation}
denotes the Schwarzschild radius of the  (lensing) compact object of
mass $M$, which evaluates to
\begin{equation}
\label{re} R_\rmn{E} = 1.73\times
10^4\left(\frac{R_\rmn{S}}{1~\mbox{km}}\right)^{1/2}\left(\frac{a}{1~\mbox{au}}\right)^{1/2}\;
\mbox{km}\,.
\label{eq:EinsteinNumerical}
\end{equation}
Given that the difference between lens and source distance as
compared to their distance from the observer can comfortably be
neglected, the Einstein radius becomes a function solely of the lens
mass and the orbital radius of the binary system, which means that
the observed signature does not depend on its distance from the
observer.

With a compact object as lens, we should however be aware of several
possible corrections to standard gravitational microlensing light
curves: {(a)} the strong gravitational field of the lensing compact
object leads to relativistic images, {(b)} geometrical corrections
due to strong fields, {(c)} the perturbation effect of the source on
the light deflection and ({d}) the finite-size effect of the source
star.

For a black hole, light rays can enter regions with strong
gravitational fields near the Schwarzschild radius and reach the
observer after a quite complicated track \citep{chand92}. Such light
rays correspond to relativistic images that exist in addition to the
usual weak-field images, and in principle affect the total
magnification pattern of the observed source star. For these
relativistic images, the relation between the source, image and
deflection angle do not satisfy the small-angle approximation, but
the lens equation for this configuration is rather given by
\begin{equation}
\tan\beta=\tan\theta-\frac{D_\rmn{LS}}{D_\rmn{S}}[\tan\theta+\tan(\alpha-\theta)],
\end{equation}
where $\theta$ and $\beta$ are the position angles of image and source,
respectively, and $\alpha$ is the deflection angle.
Integration over the path yields the deflection angle as
\begin{equation}
\alpha(x_{0})=\int^{\infty}_{x_{0}}\frac{2\,\rmn{d}x}{x\sqrt{(\frac{x}{x_{0}})^2(1-\frac{1}{x_{0}})-(1-\frac{1}{x})}}-\upi\,,
\label{diff}
\end{equation}
where all distances are in units of the Schwarzschild radius
$R_\mathrm{S}$ and $x_0$ marks the closest approach of the light ray
to the deflector. If observer, lens, and source happen to fall
exactly onto a straight line, the condition for the observation of
the source essentially becomes $\alpha = 2\upi n$, where $n$ is the
number of turning of the light rays around the black hole
\citep{Bozza01}. For source-lens (line-of-sight projected)
separations substantially larger than the Schwarzschild radius, the
magnification of the source star due to strong lensing can be
neglected as compared to the weak-field images. In this case, the
deflection angle is in the order of $\alpha\simeq
R_\rmn{E}/a\simeq\sqrt{R_\rmn{S}/a}$. With the Schwarzschild radius
$R_\mathrm{S}$ to be of the order of kilometers and the orbital
radius of the order of $10^8~\mbox{km}$, the corresponding angles in
the lens equation are in the order of $\sim 10^{-4}$, and we find
ourselves in the small-angle regime.

The proximity of the source star to the lens may also perturb the gravitational lensing effect.  Considering a
linear perturbation around the Schwarzschild metric in the weak-field limit, the perturbation on
the deflection angle relate to the Newtonian potentials as
\begin{equation}
\frac{\delta\alpha}{\alpha} = \frac{\Phi_\rmn{S}}{\Phi_\rmn{L}}\,,
\end{equation}
where $\Phi_\rmn{S}$ and $\Phi_\rmn{L}$ are the Newtonian
gravitational potentials of the source star and the lens, respectively.
For a light ray passing near the Einstein radius $R_\rmn{E}$, and source
and lens object being separated by about an
astronomical unit, one finds a relative perturbation on the deflection angle of
\begin{equation}
\frac{\delta\alpha}{\alpha}
\simeq\frac{m_\star}{M}\;\frac{R_\rmn{E}}{1~\mbox{au}}, \label{pert}
\end{equation}
where $m_\star$ and $M$ are the mass of source star and the lens, respectively.
With Eq.~(\ref{re}) one finds a numerical value of $\sim\,10^{-4}$, so that
the perturbation effect of the companion star does not play a
significant role.

Finally we look at the influence of the finite size of the observed
source star, which was discussed in detail by \citet{WM94}. The
relevant parameter $\rho_\star$ is the ratio  between the angular
radius of the source star and the angular Einstein radius, which
simplifies to $\rho_\star = R_\star/R_\rmn{E}$, given that lens and
source distances practically coincide. Eliminating the stellar
radius in favour of the stellar mass, using
$R_\star/R_\odot\simeq(m_\star/M_\odot)^{0.8}$ \citep{rmr} and using
Eq.~(\ref{eq:EinsteinNumerical}), one finds
\begin{equation}
\rho_\star = 22.7\;\left(\frac{m_\star}{M_\odot}\right)^{0.8}
\left(\frac{M}{M_{\odot}}\right)^{-1/2}\left(\frac{a}{1~\mbox{au}}\right)^{-1/2}.
\end{equation}
Given that the magnification is limited to
\begin{equation}
\mu_\rmn{max} = \sqrt{1+\frac{4}{\rho_\star^2}}\,,
\end{equation}
which is realised for perfect alignment, the signal amplitude is
quite substantially suppressed due to the finite size of
main-sequence source stars, unless the star is of low mass and/or
the compact object is a massive black hole. As pointed out by
\citet{Mae73}, white dwarfs come with a clear advantage of smaller
radii, so that larger magnifications occur regularly.

For general separations between lens and source stars, where
$u$ denotes the angular separation in units of the angular Einstein radius,
the magnification for $u\neq \rho_\star$ is given
by
\begin{eqnarray}
A(u,\rho_\star)=\frac{1}{2\upi}\Bigg[\frac{u+\rho_\star}{\rho_\star~^2}
\sqrt{4+(u-\rho_\star)^2}~~E(k)\\-\frac{u-\rho_\star}{\rho_\star~^2}~\frac{8+u^2-\rho_\star~^2}{\sqrt{4+(u-\rho_\star)^2}}~~K(k)\\+
\frac{4(u-\rho_\star)^2}{\rho_\star~^2(u+\rho_\star)}~\frac{1+\rho_\star~^2}{\sqrt{4+(u-\rho_\star)^2}}~\Pi(n;k)\Bigg],
\end{eqnarray}
where $E(k)$,$K(k$ and $\Pi(n;k)$ are the complete elliptic integral
of first , second and third kinds respectively and
\begin{equation}
 n =
\frac{4u\rho_\star}{(u+\rho_\star)^2}~~~~~~~~~~~k=\sqrt{\frac{4n}{4+(u-\rho_\star)^2}}\,,
\end{equation}
whereas for $u=\rho_\star$, one finds
\citep{Mae73,Do:thesis}
\begin{equation}
A(\rho_\star;\rho_\star)=\frac{2}{\upi}\left[\left(1+\frac{1}{\rho_\star^2}\right)\,\arcsin\frac{1}{\sqrt{1+\frac{1}{\rho_\star^2}}}
+\frac{1}{\rho_\star}\right]\,.
\end{equation}

The centre of the source star is within the angular Einstein radius
of the lens star for angles $\varphi \leq  \varphi_\rmn{max} =
R_\rmn{E}/a$. Therefore, this condition can be used as a reference
for the magnification to be substantial. We note that the
characteristic inclination angle $\varphi_\rmn{max}$ is independent
of the distance of the binary system to the observer. We find an
order estimate for the fraction of the binary systems with
significant magnification signature in their light curves as $f =
2\,\varphi_\rmn{max}/\upi$. We further find $f \sim
(2/\upi)\,(R_\rmn{E}/a)= (2/\upi)\, \sqrt{2 R_\rmn{S}/a}$. Using the
numerical values for the Schwarzschild radius in the order of a few
km and $a$ in the order of one tenth of astronomical unit, the
fraction of self-lensing binaries with compact objects that provide
a signature becomes $f \sim 10^{-4}$.
Taking 0.4 per cent of binary stars with compact star companions,
the probability for the effect to show up amongst all observed stars
turns out to be $f_\rmn{all} \sim 4 \times 10^{-7}$. This number is
tiny, but one needs to be aware of the fact that the prospects for
observing such an effect crucially depend on the viability of
regular monitoring of a huge number of targets, as well as on the
frequency of such events to occur.

For a binary system, the angular velocity is given by
\begin{equation}
\omega = \sqrt{\frac{G\,(m_\star + M)}{a^3}}\,,
\end{equation}
so that the relative transverse velocity of the source with respect to
the lens follows as
\begin{equation}
v_\perp = \omega\,a = \sqrt{\frac{G\,(m_\star + M)}{a}}\,,
\end{equation}
and is therefore determined
with the choices of the masses $m_\star$ and $M$ of the components and
the orbital radius $a$.
This defines an event time-scale
\begin{equation}
t_\rmn{E} \equiv R_\rmn{E}/v_\perp = \frac{2a}{c}\;\sqrt{\frac{M}{m_\star+M}}\,,
\label{eq:tE}
\end{equation}
within which the source moves by $R_\rmn{E}$. In fact, the motion can be approximated as uniform, where
\begin{equation}
u(t) = \sqrt{u_0^2 + \left(\frac{t-t_0}{t_\rmn{E}}\right)^2}\,,
\end{equation}
with the closest angular approach between lens and source star being
\begin{equation}
u_0 = \frac{a}{R_\rmn{E}}\;\varphi = \sqrt{\frac{a}{2\,R_\rmn{S}}}\;\varphi
\label{eq:u0}
\end{equation}
for a small $\varphi$, which occurs at epoch $t_0$. Therefore, the signal of eclipsing microlensing
resembles an normal extended-source standard microlensing light curve,
described by the 4 parameters $t_\rmn{E}$, $t_0$, $u_0$, and $\rho_\star$.

\begin{figure}
\begin{center}
\includegraphics[height=5.2cm,width=7.8cm]{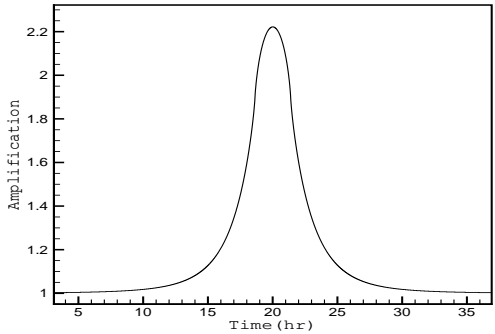}
\caption{\label{fig2} Gravitational self-microlensing light curve
arising from a binary system that involves a black-hole lens of mass
$M = 8.5~M_\odot$ and an observed main-sequence star of mass
$m_\star = 0.35~M_\odot$. The orbit is $\varphi = 0.33\arcsec$ from
an edge-on configuration, and the orbital radius is $a=
17~\mbox{au}$. This yields a finite-size parameter $\rho_\star =
0.81$ and an orbital period $P \sim 23~\mbox{yrs}$.}
\end{center}
\end{figure}

For reference, the light curve of a binary system with the
parameters of $M = 8.5~M_\odot$, a main sequence star with the mass
of $m_\star = 0.35~M_\odot$, $a= 17~\mbox{au}$ and $\varphi =
0.33\arcsec$ is shown in Fig.~\ref{fig2}. This system has the
finite-size parameter $\rho_\star = 0.81$ and the period of this
system is about $23$ years. Main-sequence stars are again
disfavoured due to their long periods in detectable systems, whereas
substantial signals can arise in systems with white dwarfs with much
shorter periods.

\section{Detection probability}
\label{montecarlo}

Let us now investigate the prospects for detecting compact objects
by means of binary self-lensing for specific observational
strategies. Modelled upon the characteristics of current or upcoming
microlensing campaigns, and giving us a hint on the roles of both
photometric accuracy and sampling rate, we consider regular
monitoring with the following parameters \citep[see also][]{rah}:
(a) 5 per cent photometric accuracy at 15~min cadence, indicative
for high-cadence ground-based surveys \citep{Sumi:planet,KMTNet},
(b) 2 per cent accuracy at 2~hr cadence, roughly representative of
current follow-up monitoring programmes \citep{PLANET:EGS}, and (c)
0.3 per cent photometric accuracy at 15-min cadence, reflecting the
coming state-of-the-art, including lucky-imaging or spaced-based
observations \citep{Uffe:planets,Bennett:space1,Bennett:space2}.

For main-sequence stars, we adopt the mass function $\xi(m_\star) =
dN/d[\lg (m_\star/M_\odot)]$ proposed by \citet{chab03}, namely
\begin{equation}
\xi(m_\star) = \left\{\begin{array}{l}
0.093\exp\left\{-\frac{[\lg(m_\star/M_\odot) -
\lg(0.2)]^2}{2\times(0.55)^2}\right\}  \\
\hfill \mbox{for} \quad
 m_\star<1~M_\odot\,, \\[2ex]
0.041~(m_\star/M_\odot)^{-1.35} \\
\hfill \mbox{for} \quad m_\star \geq 1~M_\odot\,.
\end{array}
\right.\,,
\label{mf}
\end{equation}
which covers the range
of $m_\star\in[0.1,2]~ M_\odot$, while we assume a mass-radius
relation $R_\star/R_\odot\simeq(m_\star/M_\odot)^{0.8}$ \citep{rmr}.

For the compact objects, we adopt the product of the evolution of
the zero-age mass function to the final stage of stars \citep{bel02}
with the mass range of $M\in[1.2,15]~M_\odot$. To estimate the
fraction of binary systems with one compact object and one main
sequence star, we do a rough calculation for stars in the binaries
with the initial masses in the range of $M<1 M_\odot$ for the first
star and $M> 8M_\odot$ for the companion star. Star with the larger
mass has a relative short life time and will evolve to a compact
object while the smaller star stays in the main sequence if we don't
have mass transfer between the two stars. For the binaries located
far enough distance from each other (i.e. stellar size should be
smaller than the roche lobe), we obtain almost $0.4$ per cent of the
stars will end to the binary systems with one compact object and a
companion main sequence star.

For the orbital distance within the binary system, we assume a
logarithmic distribution in the range of $a\in [0.01,50]~\mbox{au}$,
in accordance with \"{O}pik's law, while the inclination angle is
drawn uniformly from $\varphi \in [0,\upi/2]$.
\begin{figure}
\begin{center}
\includegraphics[height=6.5cm,width=7.5cm]{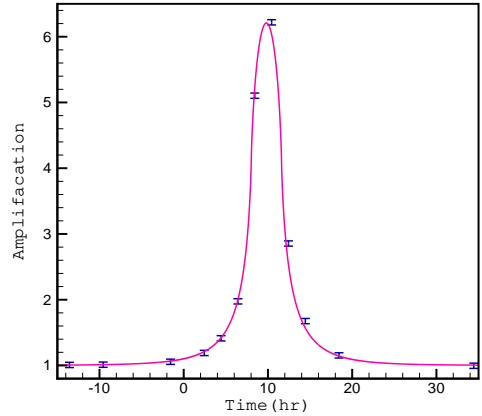}
\caption{\label{lc_sim} Example synthetic light curve as arising
from the Monte-Carlo simulation. The adopted parameters are $M =
13.28~M_\odot$ $m_\star = 0.2~M_\odot$, $d=30~\mbox{au}$, and
$\varphi = 0.01\arcmin$, so that
$\rho_\star=0.32$ and $t_\rmn{E} = 5.95~h$.}
\end{center}
\end{figure}

Using these parameter distributions, we generated synthetic light
curves by means of Monte-Carlo simulations, where
Figure~\ref{lc_sim} shows an example. With a detection criterion of
three consecutive data point being larger than three times of the
standard deviation from the base line, we not only obtain the
fraction of systems for which the compact object is detectable, but
also the distribution of parameters of the expected eclipsing
microlensing events.

\begin{figure}
\begin{center}
\includegraphics[height=8.cm,width=9.5cm]{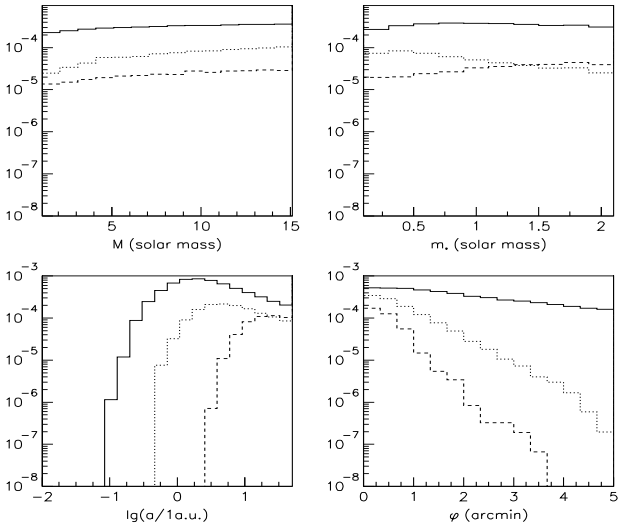}
\caption{\label{effms} The efficiency $\varepsilon$ for revealing
the presence of a compact object in a binary system with an observed
main-sequence star as a function of $M$ (lens mass), $m_\star$
(source mass), $a$ (orbital radius), and $\varphi$ (inclination
angle) for three observational setups, characterised by their
photometric accuracy $\sigma$ and sampling interval $\Delta t$
(dotted: $\sigma = 5~\mbox{per cent}$, $\Delta t = 15~\mbox{min}$;
dashed: $\sigma = 2~\mbox{per cent}$, $\Delta t = 120~\mbox{min}$;
solid: $\sigma = 0.3~\mbox{per cent}$, $\Delta t = 15~\mbox{min}$).}
\end{center}
\end{figure}

Figure~\ref{effms} shows
the detection efficiency for the three considered monitoring strategies.
One finds that it depends only weakly on the mass of
the lens. This is a consequence of the relation between the lens mass
$M$ and the event time-scale $t_\rmn{E} = R_\rmn{E}/v$.
With $R_\rmn{E} \propto \sqrt{M}$ and $v_\perp \propto \sqrt{m_\star+M}$,
one finds a weakly-varying $t_\rmn{E} \propto \sqrt{{M}/{(M + m_\star})}$. A larger mass $m_\star$ of the main-sequence source star implies
a larger radius $R_\star$, which diminishes the magnification due to the
finite-size effect. Moreover, the event time-scale becomes smaller.
On the other hand, a larger source radius $R_\star$ enables us to get
a signal from a wider range of inclination angles, and the effective signal duration is increased. The gain from a longer signal duration plays a
larger role for sparser sampling, while for an inferior photometry
the signal drops below the detection threshold earlier.

The effect of the orbital radius of the two companion stars on the
observability eclipsing microlensing signal is a function of three
factors, namely (a) the dependence of the Einstein radius on the orbital
radius as $R_\rmn{E}\propto\sqrt{a}$, (b) the relative transverse
velocity of the binary system $v\propto 1/\sqrt{a}$, hence
$t_\rmn{E}\propto a$, and (c)
$\varphi_\rmn{max} = R_\rmn{E}/a \propto 1/\sqrt{a}$. The wider range of suitable inclination
angles increases the prospects for a detection in systems with smaller
orbital radius. Smaller event time-scales however let signals fall
into the gap between subsequent observations. Consequently,
we find a rise in the detection efficiency towards smaller orbital radii
(and thereby shorter periods) until the signals become to short to be detectable.

\begin{figure}
\begin{center}
\includegraphics[height=7.5cm,width=8.5cm]{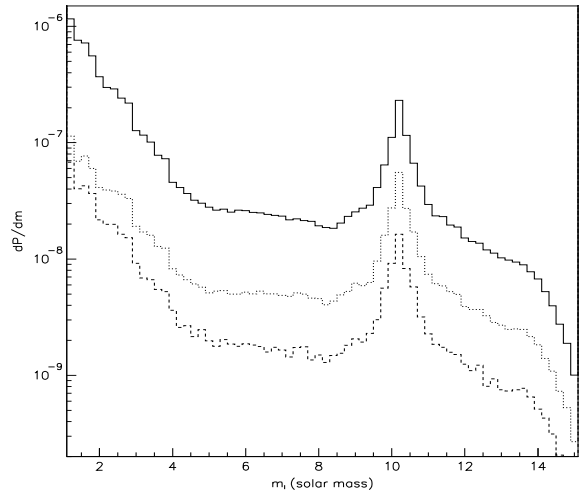}
\caption{\label{prob1} Expected distribution of the masses of the
detected compact objects that act as gravitational (micro)lenses on
the light of observed main-sequence star within a binary system,
considering the same observational capabilities as for
Fig.~\protect\ref{effms}.}
\end{center}
\end{figure}

With the detection efficiency and the distribution functions of the
adopted parameters, we find the overall probability for detecting
binary self-microlensing events. In particular, by multiplying the
detection efficiency with the mass function of the lens stars, we
obtain the expected distribution of lens masses revealed from
observed eclipsing microlensing signals, which is shown in
Fig.~\ref{prob1}. The mass function of the lens stars were
normalized to the overall number of stars. Integrating these
histograms results in the total probability of observing eclipsing
microlensing events. For our three variants of the adopted observing
strategy, we find $f_\rmn{all} = 1.45\times 10^{-7}$,
$f_\rmn{all}=6.50\times 10^{-8}$, or $f_\rmn{all}=9.97\times
10^{-7}$ respectively.
With the latter value being close to our earlier thumb estimate, we
find a rather good efficiency of the adopted strategy.

We further weigh each detection efficiency $\varepsilon$ with the
frequency of the signal, which equals the inverse of the orbital
period $P$, i.e.\ we calculate an average
$\left<\varepsilon/P\right>$ over the realisations arising from the
Monte-Carlo simulation, in order to obtain the event rate per
observed star as $\gamma = 3.71 \times 10^{-7}~\mbox{yr}^{-1}$,
$\gamma = 2.08 \times 10^{-8}~\mbox{yr}^{-1}$, or $\gamma =
3.28\times 10^{-6}~\mbox{yr}^{-1}$ for our three adopted monitoring
strategies, which typically find compact objects in binaries with
orbital periods of $P \sim 0.39~\mbox{yr}$, $P \sim 3.12~\mbox{yr}$,
or $P \sim 0.30~\mbox{yr}$ respectively, which equals the period of
recurrence of the signals. Naturally, systems with shorter periods
dominate the events due to their higher recurrence rate, and the
goal of an observational strategy has to be to keep these
detectable.
 The findings of our simulations are summarized in Table~\ref{tab:values}.

\begin{table}
\begin{tabular}{ccccc}
\hline
accuracy & sampling rate & detectability & event rate & period \\
$\sigma$ & $\Delta t$ & $f_\rmn{all}$ & $\gamma$ & $\hat P$ \\
$[\mbox{per cent}]$ & $[\mbox{min}]$ &  & $[\mbox{yr}^{-1}]$ & $[\mbox{yr}] $
\\
\hline
$5$ & $15$ & $1.45 \times 10^{-7}$ & $3.71 \times 10^{-7}$ & $0.39$ \\
$2$ & $120$ & $6.50 \times 10^{-8}$ & $2.08 \times 10^{-8}$ & $3.12$ \\
$0.3$ & $15$ & $9.97 \times 10^{-7}$ & $3.28 \times 10^{-6}$ & $0.30$ \\
\hline
\end{tabular}
\caption{Fraction of observed systems with a detectable compact companion
$f_\rmn{all} = \left<\varepsilon\right>$, event rate per observed system
$\gamma = \left<\varepsilon/P\right>$, and 'typical' period
$\hat{P} = \left<\varepsilon\right>/\left<\varepsilon/P\right>$ of the signal for the three considered monitoring strategies characterized
by the photometric accuracy $\sigma$ and the sampling interval
$\Delta t$,
where $\varepsilon$ denotes the detection efficiency for a given configuration, and $P$ denotes its orbital period.}
\label{tab:values}
\end{table}

\section{Extraction of parameters}
\label{deg}
The observed light curve allows to extract the 4 standard parameters $t_0$,
$u_0$, $t_\rmn{E}$, and $\rho_\star$, but with $t_0$ not carrying any
relevant information about the binary system, we are one parameter short
of reconstructing the masses of the components $m_\star$ and $M$, the
orbital radius $a$, and the inclination angle $\varphi$.
Only in the limit $m_\star \ll M$, Eq.~(\ref{eq:tE}) yields
\begin{equation}
a = \frac{c\,t_\mathrm{E}}{2}\,\sqrt{\frac{M + m_\star}{M}}
\simeq \frac{c\,t_\mathrm{E}}{2}\,.
\end{equation}

In order to go further, one needs to exploit the periodicity of the
signal. This again stresses the need for events with shorter
periods, not longer than a few years. In fact, any attempt to obtain
information by measuring astrometric shifts of the observed source
star due to its wobble around the compact object or its radial
velocity by means of Doppler-shifts of spectral lines, relates to
the orbital period. Withstanding the difficulties in obtaining such
measurements for faint stars, the fundamental properties already
follow with the orbital period itself.

Kepler's third law
\begin{equation}
P = 2\upi\;\sqrt{\frac{a^3}{G\,(M+m_\star)}}
\end{equation}
would allow to find
\begin{equation}
\varphi = 2\upi\,u_0\,\frac{t_\rmn{E}}{P}
\end{equation}
with Eqs.~(\ref{eq:tE}) and~(\ref{eq:u0}),
and one would be able to obtain iteratively
\begin{equation}
M = \frac{4 \pi^2}{G P^2}\;a^3 - m_\star \simeq \frac{\pi^2 c^3 t_\rmn{E}^3}{2 G P^2}\,,
\end{equation}
as well as
\begin{equation}
R_\star = \frac{2\rho_\star}{c}\;\sqrt{GMa} \simeq \frac{\upi\,\rho_\star\,c\,t_\rmn{E}^2}{P}\,,
\end{equation}
so that with the mass-radius relation for main-sequence stars
\begin{equation}
m_\star = M_\odot\,\left(\frac{R_\star}{R_\odot}\right)^{5/4}
       \simeq M_\odot\,\left(\frac{\upi\,\rho_\star\,c\,t_\rmn{E}^2}{P R_{\odot}}\right)^{5/4}\,.
\end{equation}

\section{Conclusions}
\label{conclusion}

Given that the signal amplitude of self-lensing due to a compact
object in a binary system is less suppressed by the much smaller
finite radius of a white dwarf as compared to a main-sequence star,
and moreover the orbital period of detectable systems is smaller
(given that the relevance of finite-source effects is quantified by
$\rho_\star \propto 1/\sqrt{a}$), and thereby the frequency of
signals is larger, \citet{Mae73} concluded that white dwarfs are the
favourable targets for observing this effect, whereas the prospects
for binaries involving main-sequence stars are rather bleak.
However, the fortune changes substantially if one looks at the
observability of suitable systems. \cite{BT} considered the Sloan
Digital Sky Survey (SDSS) as most favourable for observing white
dwarfs, and in fact, it has dramatically increased the number of
known white dwarfs. However, with the sample containing about 15,000
objects \citep{SDSS}, it is $\sim\,10^{4}$ times smaller as compared
to the $2 \times 10^{8}$ stars regularly monitored by current
microlensing surveys \citep{OGLE-III}.

For $N_\rmn{obs} \sim 2 \times 10^{8}$ monitored stars and an event
rate per observed star of $\gamma \sim 4 \times
10^{-7}~\mbox{yr}^{-1}$ (for 5 per cent photometric accuracy and
15~min sampling cadence), one finds a total event rate of $\Gamma
\sim 74\,\kappa~\mbox{yr}^{-1}$, where $\kappa < 1$ is a coverage
factor accounting for the visibility of the Galactic bulge from the
respective sites over the year, any losses due to weather or
technical downtime, and imperfect cadence or data quality. In
contrast to earlier work, we therefore conclude that the detection
of compact objects (in fact, predominantly black holes) in binary
systems due to self-lensing of an observed main-sequence star
companion is possible, provided that a high-cadence sampling
substantially below 2~hrs is realised. The upcoming Korea
Microlensing Telescope Network (KMTNet) has in fact been designed as
a wide-field survey of the Galactic Bulge with 10-minute cadence
\citep{KMTNet}. Moreover, the MOA (Microlensing Observations in
Astrophysics) survey already monitors some of its fields at that
cadence \citep{Sumi:planet}. Higher photometric accuracies of 0.3
per cent, achievable with space-based observations
\citep{Bennett:space1,Bennett:space2} or lucky-imaging cameras
\citep{Uffe:planets}, could result in 10 times as many observable
signals due to self-lensing in binaries with a compact objects, whereas lower accuracies of 20 per cent would lead to about 10 times less objects being detected.

Given that the duration of the expected self-microlensing signals is
of the order of a few hours, we issue a note of caution that such is
not mistaken for evidence of planetary-mass bodies that pass the
line of sight to a background star. In fact, the MOA survey appears
to show an excess of short-duration peaks as compared to
expectations from stellar populations and the kinematics of the
Milky Way (K.~Kamiya, private communication).

In practice, one faces a rather hard job to distinguish between
usually poorly-covered spikes of different origin. The self-lensing
binary signals repeat in principle, but on an initially unknown
time-scale of months to years and are rather easy to miss. The
discriminating power of the criterion of achromaticity of
gravitational microlensing as opposed to stellar variability is also
limited due to the lack of detail on the shape of the signal. Only
if a period of the binary system can be established, its physical
characteristics can be determined.

\section*{Acknowledgments}
We would like to thank Valerio Bozza for a couple of helpful remarks on this topic.

\bibliographystyle{mn2e}

\bibliography{compactbinary}

\end{document}